\documentclass{article}
\usepackage{spconf,amsmath,graphicx, tabularx, siunitx} % Required for inserting images
\usepackage[T1]{fontenc}
\usepackage{caption} 
\title{A lightweight dual-stage framework for personalized speech enhancement based on DeepFilterNet2}
\name{Thomas Serre$^{\star \dagger}$, 
    Mathieu Fontaine$^{\star}$, 
    Éric Benhaim$^{\dagger}$, 
    Geoffroy Dutour$^{\dagger}$, 
    Slim Essid$^{\star}$ }

\address{$^{\dagger}$ Orosound, Signal Processing Lab \quad 
        $^{\star}$LTCI, Télécom Paris, Institut Polytechnique de Paris}

\begin{document}

\maketitle

\begin{abstract}

Isolating the desired speaker's voice amidst multiple speakers in a noisy acoustic context is a challenging task. Personalized speech enhancement (PSE) endeavours to achieve this by leveraging prior knowledge of the speaker's voice.
Recent research efforts have yielded promising PSE models, albeit often accompanied by computationally intensive architectures, unsuitable for resource-constrained embedded devices.
In this paper, we introduce a novel method to personalize a lightweight dual-stage Speech Enhancement (SE) model and implement it within DeepFilterNet2, a SE model renowned for its state-of-the-art performance.
We seek an optimal integration of speaker information within the model, exploring different positions for the integration of the speaker embeddings within the dual-stage enhancement architecture.
We also investigate a tailored training strategy when adapting DeepFilterNet2 to a PSE task. 
We show that our personalization method greatly improves the performances of DeepFilterNet2 while preserving minimal computational overhead.
    
\end{abstract}


\section{Introduction}
% SE --> PSE
%Just like humans can focus on one voice in a noisy environment when they have already heard it in the past, 
Personalized speech enhancement (PSE) aims to extract the voice of a speaker with prior information on their voice. 
Such systems become very interesting when it comes to making a call in a crowded environment, working from home or even enhancing the voice of hearing aid users. 

Over the years, deep neural network (DNN) based speech enhancement frameworks have emerged and achieved superior results, even in particularly noisy environments. 
However, such architectures usually perform poorly in an environment where interfering voices overlap with the voice of interest.
PSE is therefore well suited for this task,  as it benefits from prior information on the target speaker. 
This information is usually extracted from a speaker-specific enrollment clip using an encoder model. Then, it is fed to a downstream speech enhancement model and used as a cue to identify the target voice in the noisy content. 
These two models can either be trained jointly \cite{xu2020spex, ge2020spex+, ge2021spex++, ji2020speaker_joint} or separately \cite{wang2018voicefilter, giri2021persopercepnet, newmodels_pdccrn}. 
Frameworks like VoiceFilter \cite{wang2018voicefilter}, SpEx \cite{xu2020spex}, SpeakerBeam \cite{vzmolikova2019speakerbeam} or pPercepNet \cite{giri2021persopercepnet} paved the way for the PSE task showing its superiority to standard SE, especially with interfering voices. 

% Multi stage SE --> MultiStage PSE
Recently, the multi stage approach has been introduced to PSE after showing great results in speech enhancement \cite{2stage}.
The task is thus no longer performed as one task, but it is divided into several sub tasks instead. 
Ju, Yukai, et al. showed the benefits of the multistage approach in PSE by introducing the TEA-PSE \cite{ju2022tea} composed of a sub network called the MAG-Net, that estimates the clean magnitude, and another, the COM-Net, that estimates the clean complex spectrogram. 

% Real time PSE
Although the multistage approach has led to smaller and more effective architectures in standard SE, such as GaGNet \cite{li2022glance} or DeepFilterNet2 \cite{schroter2022deepfilternet2}, recent dual-stage models in PSE tend to be much larger.
Many studies have worked on lighter PSE models that can be used in real-time on embedded devices \cite{thakker2022fast, wang2020voicefilter, delcroix2019compact}. However,  the approach of adapting a lightweight dual-stage SE framework to PSE has not been considered yet, despite being critical for PSE on embedded devices, where system complexity must be controlled.

% Contribution
In this work, we propose a method to adapt a dual stage state-of-the-art SE model to the PSE task. 
We apply this method to DeepFilterNet2, a lightweight dual-stage framework renowned for its performance and its compact architecture. 
We seek an optimal integration of speaker information within the model, by exploring different positions for the integration of the speaker embeddings within the dual-stage enhancement architecture. 
% We also design a tailored dataset and show the importance of the plurality of interfering speakers
% We also investigate a tailored training strategy when adapting DeepFilterNet2 to a PSE task. 
We demonstrate that our personalization approach significantly enhances the performance of DeepFilterNet2\footnote{Audio demo available at: http://pdeepfilternet2.github.io/} while maintaining a negligible increase in computational resource utilization.

\section{Proposed method}

\subsection{Speaker Encoder: ECAPA-TDNN}

The speaker encoder is used to encode the acoustic information of the target speaker into an embedding. 
Given an observation $x_{obs}$, the speaker encoder transforms it into a vector $x_e \in R^D$, where $D$ is the dimension of the latent space of the embedding. 
This process is summarized as $x_e = E(x_{obs})$ where $E$ corresponds to the speaker encoder, and $x_e$ to the embedding.
In this work, we use the ECAPA-TDNN \cite{ecapa} as the speaker encoder. 
This framework achieves state-of-the-art results in speaker recognition and is widely used in PSE. 
Greatly inspired by the X-vector \cite{snyder2018xvector}, it consists of Time Delay Neural Network (TDNN) layers followed by an attention pooling layer. % and is trained using the AMM-softmax loss. 
In our case, the encoder is trained before the enhancement network training, and its weights are frozen so that they remain unchanged during the training of the enhancement model.

\subsection{Enhancement Network: DeepFilterNet2}

\subsubsection{DeepFilterNet2}
\label{sec:deepfiltenet2}
DeepFilterNet2 \cite{schroter2022deepfilternet2} is a very light framework that achieves state-of-the-art results in speech enhancement. 
Based on a U-Net type of architecture, this framework is composed of two separated blocks. 
The first block performs a coarse estimation using Equivalent Rectangular Bandwidth (ERB) features as input, and the second block performs a finer estimation using complex domain spectrograms and Deep Filtering (DF). 
The latter is applied to the output of the coarse estimation, thus instantiating the dual-stage framework. 
In this work, we mainly focus on the encoder, which can be summarised as follows:
\begin{equation}
    X_{enc}(k) = F_{enc}(X_{erb}(k, b) , X_{df}(k,f_{df})) \\
\end{equation}

where $X_{erb}$ is the ERB feature, $X_{df}$ the complex feature and $F_{enc}$ the encoder of DeepFilterNet2. Then, the first stage estimates only the speech envelope and the second stage performs Deep Filtering on the output of stage 1. The filter coefficients are generated from the lower part of the complex spectrogram and $f_{df}=5kHz$.

\subsubsection{Personalized model: pDeepFilterNet2}
\label{section: pDeepFilternet}

\begin{figure}[t]
    \centering
    \includegraphics[width=0.95\linewidth]{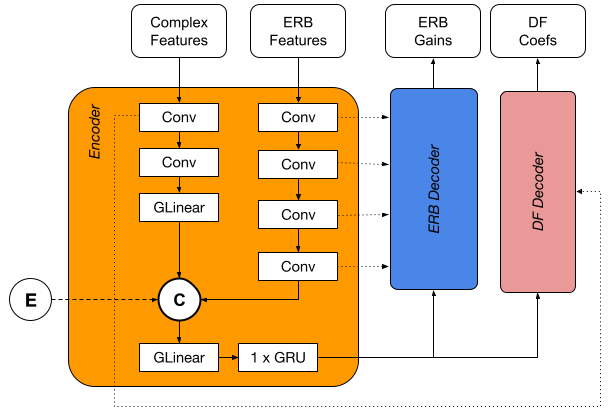}
    \includegraphics[width=0.95\linewidth]{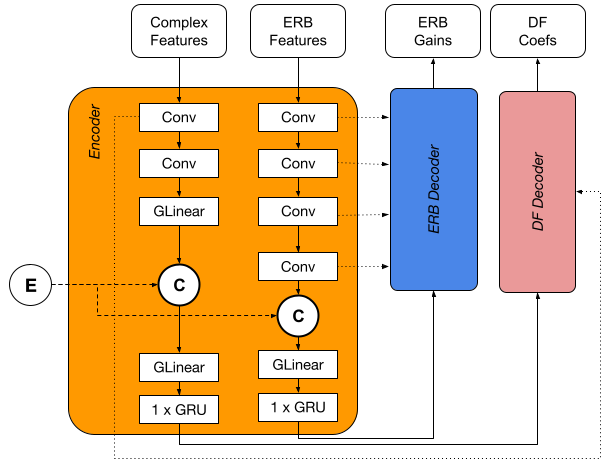}
    \caption{Personalized DeepFilterNet2 with unified encoder (top) - Personalized DeepFilterNet2 with dual encoder (bottom) - \textbf{\fontfamily{phv}\selectfont E} represents the embedding and \textbf{\fontfamily{phv}\selectfont C} is the concatenation operation.}
    \label{fig:perso_v1}
\end{figure}

We propose pDeepFilterNet2, the personalized version of DeepFilterNet2 \cite{schroter2022deepfilternet2}. 
The personalization of this framework is done by introducing the embedding of the pre-trained ECAPA-TDNN into the speech enhancement model. 
We thus investigate several locations to add the embedding. 
As most of the feature extraction and analysis occurs within the encoder, we mainly focus on the encoder.
More precisely, we propose two architectures: the unified encoder and the dual encoder (see Fig. \ref{fig:perso_v1}).

\noindent \textbf{Unified encoder.} Our first intuition is to simply concatenate the embedding where the two branches of the encoder are joined, as shown on the top schematics of Fig. \ref{fig:perso_v1}. This is a straightforward approach that allows the speaker information to be used by both branches while minimizing the computational cost. The encoder can thus be modelled as follows.
\begin{equation}
    X_{enc}(k) = F_{enc}(X_{erb}(k, b) , X_{df}(k,f_{df}), x_e) \\
\end{equation}

where $x_e$ corresponds to the embedding.
However, this approach does not allow us to understand  what branch benefits the most from the embedding. Therefore, we also propose a dual encoder version to better leverage the embeddings.
%dive more into the details of the embeddings added value. 

\noindent \textbf{Dual encoder.} 
On the bottom schematics of Fig. \ref{fig:perso_v1}, we propose the dual encoder version of the personalized DeepFilterNet2, we call it pDeepFilterNet2$^D$. Splitting the encoder into two independent branches allows us to get a fair comparison between adding the embeddings in the ERB branch, or in the DF branch. That way, we can understand how the embedding is used and where its addition is the most effective. The dual encoder is modelled as follows. 
\small
\begin{equation}
    X_{ERB}(k)=F_{ERB\_enc}(X_{erb}(k, b), x_e)
\end{equation}
\begin{equation}
    X_{DF}(k)=F_{DF\_enc}(X_{df}(k,f_{df}), x_e) 
\end{equation}
\normalsize

Thus, this architecture allows us to test three possibilities: adding the embedding in both branches, in the ERB branch only, and in the DF branch only. These three models are respectively named as follows: pDeepFilterNet$^D_{both}$, pDeepFilterNet$^D_{erb}$ and pDeepFilterNet$^D_{df}$.

\subsubsection{Loss function}

The training objective remains almost the same as the one used for the training of DeepFilterNet2. It consists of a spectral loss $\mathcal{L}_{spec}$, a multi-resolution loss $\mathcal{L}_{MR}$ and an over-suppression loss $\mathcal{L}_ {OS}$ \cite{osloss}.
$\mathcal{L}_{MR}$ is computed for the 4 following windows \{5,10,20,40\} ms. The overall loss can thus be summarized as follows.
\begin{equation}
    \mathcal{L} = \lambda_{spec}\mathcal{L}_{spec} + \lambda_{MR}\mathcal{L}_{MR} + \lambda_{OS}\mathcal{L}_{OS}
\end{equation}

where $\lambda_{spec}$, $\lambda_{MR}$ and $\lambda_{OS}$ correspond to loss weights which are found during training.

\section{Experiments}

\subsection{Speaker Encoder training}
In this work, we do not focus on the training of the speaker encoder. 
Therefore we simply use the pretrained ECAPA-TDNN\footnote{https://github.com/TaoRuijie/ECAPA-TDNN} and its weights are frozen during the training of pDeepFilterNet2. 
The pretrained version was trained on VoxCeleb2 and achieved 0.96\% EER on the VoxCeleb1 test set. The dimension of the embeddings is set to $D=192$.

\begin{table*}[t]
    \caption{Results on generated test set. For evaluation scores, higher is better. \textbf{Bold} means best results.}
    \centering
    {\renewcommand{\arraystretch}{1.1}%
    \begin{tabularx}{\linewidth}{l l*{7}{c} | c*{3}{c}}
    \hline
    & Model          &&& PESQ & STOI & CSIG & CBAK & COVL & Params (M) & MACs (G) & RTF   \\ 
    \hline
    
    & Noisy                && & \num{1.81416}& \num{0.74938} & \num{2.98951} & \num{2.44852} & \num{2.36840} & -      & -    & - \\
    
    & DeepFilterNet2        && & \num{2.099283}& \num{0.748479} & \num{3.114178} & \num{2.66037} & \num{2.576958} & \textbf{\num{2.31}}  & \textbf{\num{0.33}} & \textbf{\num{0.025133}}\\
    \hline
    
    & pDeepFilterNet2       && & \textbf{\num{2.362}}& \num{0.784} & \textbf{\num{3.656}} & \textbf{\num{2.898}} & \textbf{\num{3.005}} & \textbf{\num{2.31}}  & \textbf{\num{0.33}} & \textbf{\num{0.0253718}}\\
    
    & pDeepFilterNet2$^{D}_{both}$ && & \num{2.310}& \num{0.777} & \num{3.598} & \num{2.830} & \num{2.949} & \num{2.71}  & \num{0.40} & \textbf{\num{0.0256169}}\\
    
    & pDeepFilterNet2$^{D}_{erb}$  &&& \num{2.211}& \num{0.770} & \num{3.514} & \num{2.758} & \num{2.849} & \num{2.71}  & \num{0.40} & \textbf{\num{0.0256169}}\\
    
    & pDeepFilterNet2$^{D}_{df}$ && & \num{2.316}& \textbf{\num{0.787}} & \num{3.577} & \num{2.845} & \num{2.941} & \num{2.71}  & \num{0.40} & \textbf{\num{0.0256169}}\\
    \end{tabularx}} \quad
    
    \label{tab:synth}
\end{table*}

\begin{table}[t]
    \caption{PDNSMOS P.835 results on the DNS5 blind test set. Higher is better. \textbf{Bold} means better than noisy. \underline{Underline} means best results.}
    \centering
    {\renewcommand{\arraystretch}{1.3}%
    \resizebox{0.99\linewidth}{!}{
    \begin{tabularx}{1.2\linewidth}{l*{3}{c}|*{3}{c}}
    \hline
    Model & \multicolumn{3}{c}{Track 1: Headset} & \multicolumn{3}{c}{Track 2: Speakerphone}  \\\hline
          & SIG   & BAK   & OVRL        & SIG    & BAK   & OVRL     \\\hline
    Noisy &  \underline{\num{4.152}} & \num{2.369} & \num{2.709} & \underline{\num{4.0460}} & \num{2.159} & \num{2.497} \\
    Proposed&  \num{3.69}  & \textbf{\num{3.51}} & \textbf{\num{3.04}} & \num{3.63}& \textbf{\num{3.38}} & \textbf{\num{2.94}} \\ 
    TEA-PSE 3.0 & \num{4.108} & \underline{\num{4.053}} & \underline{\num{3.645}} & \num{3.993} & \underline{\num{3.951}} & \underline{\num{3.493}} \\
    \hline
    \end{tabularx}}} \quad
    
    \label{tab:dnsmos}
\end{table}

% Noisy    SIGMOS:  4.15181 | Enhanced SIGMOS:  3.44613 | Noisy    BAKMOS:  2.36081 | Enhanced BAKMOS:  3.22120 | Noisy    OVLMOS:  2.70314 | Enhanced OVLMOS:  2.70844 | Noisy    p808MOS:  3.11184 | Enhanced p808MOS:  3.06690

\subsection{pDeepFilterNet training}
\label{pdeepfiltenet_training}
\subsubsection{Dataset}
\label{dataset}

% DN5 for speech and noise
% Mozilla for interfering (and AMI)
% clean speech --> had to clean : 2448 speakers
% VCTK for test

For training set generation, we used the DNS5 personalized dataset \cite{dubey2023icassp} for target speech and noise, and we used Mozilla Common Voice (MCV) \cite{ardila2019common} for interfering speech. 
The DNS5 personalized dataset contains 3230 speakers with a total duration of 750 hours.
However, when looking at some samples of clean speech in the dataset, we found out that some recordings contained several speakers. 
This problem was mentioned in \cite{wang2023upn} and we used the same method to remove unwanted speakers. 
We reserved VCTK data for test set generation resulting in 2448 target speakers in the training set. 
We then completed our training with the MCV corpus for interfering speech. We used 261 hours with 7069 different voices. 
We generated 950 hours in total and the excerpts were split into three categories: \textit{target+noise}, \textit{target+interfering} and \textit{voice+interfering+noise}. 
The distribution of those excerpts in the dataset is respectively 20\%, 30\% and 50\%. 
The SNR and SIR are drawn from a Gaussian distribution in [-5; 35] dB and [-5, 25] dB respectively. 
Finally, we reserve around 50 hours of the generated dataset for validation.

\subsubsection{Training setup}

The training procedure is very similar to the one of DeepFilterNet2.
The model was trained with the Adam optimizer with the same learning rate and weight decay as in \cite{schroter2022deepfilternet2}. 
We also used batch scheduling from 8 to 128 to speed up the convergence. 
We perform early stopping on the validation loss with a patience of 15 epochs. 
The loss factors are set like in the original paper: $\lambda_{spec}=1e3$ and $\lambda_{MR}=5e2$, and $\lambda_{OS}=5e2$.
We also keep the same computational parameters by choosing 20-ms windows, an overlap of 50\%, and a lookahead of 2 frames resulting in a 40-ms lookahead. Finally, we also take 32 ERB bands, $f_{df}=5$ kHz, and $N=5$ for the Deep Filtering order.

\section{Results}
\label{sec:results}

We evaluate our models on two test sets. 
The first one is synthetically generated and the second one is the DNS5 blind test set. 
To generate our test set, we use the VCTK corpus for both target speech and interfering speech, and DNS5 challenge for the additive noise. 
We equally generate \textit{target+noise}, \textit{target+interfering} and \textit{voice+interfering+noise} excerpts. The SNR and the SIR are uniformly drawn in [-5, 35] dB and [-5, 25] dB respectively. 
In total, we generate 3600 files of 5 s.

\subsection{Synthetic test set}

The results on the synthetic test set are featured in table \ref{tab:synth}. 
We thus compare the four personalization methods mentioned in \ref{section: pDeepFilternet} to the non-personalized version of DeepFilterNet2 and we also compare the personalized models together.
We use the same metrics as the one used in \cite{schroter2022deepfilternet2} which are the PESQ, the STOI and the composite metric.

\begin{figure}[t]
    \centering
    \includegraphics[width=0.8\linewidth]{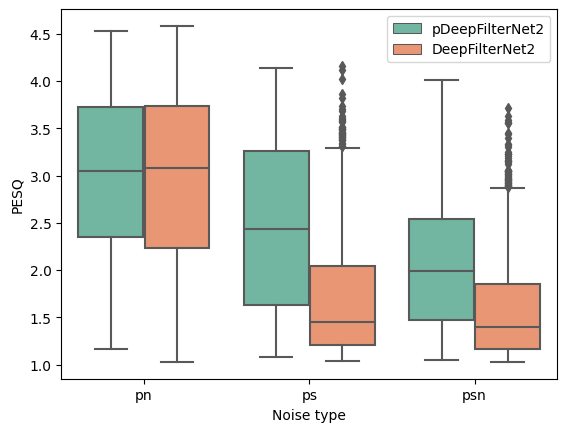}
    \caption{Box plot featuring the PESQ for DeepFilterNet2 and pDeepFilterNet2 (unified encoder version) for different noise types: primary speaker + noise (pn), primary speaker + secondary speaker (ps), primary speaker + secondary speaker + noise (psn).}
    \label{fig:boxplot}
\end{figure}

% \subsubsection{Comparison with DeepFilterNet2} 
\noindent \textbf{Comparison with DeepFilterNet2.}\ 
% What do we compare ? 
% How do we compare ? Which metrics ?
% Better with personalisation for all personalisation
% 
We observe that all personalized models perform better than DeepFilterNet2.
This improvement can especially be seen with the PESQ and the CSIG metrics. This underlines the added value of the embedding to recover the target voice. 
This can also be observed on Fig. \ref{fig:boxplot} where we compare DeepFilterNet2 and pDeepFilterNet2 using the PESQ on subsets of the test set. We see that on the \textit{pn} subset, both models achieve the same PESQ. However, on \textit{ps} and \textit{psn} subsets, the personalized model performs much better. This result was expected, and showcases the benefit of the embedding in the presence of an interfering speaker.

% \subsubsection{Embedding integration}
\noindent \textbf{Embedding integration.}\ 
We also study the position of the embedding integration. Among all personalized models, the unified encoder model is the best one. 
This suggests that concatenating the speaker embedding with the features of the two branches leads to a better use of the embedding in the network. 
However, even though pDeepFilterNet2$^{D}_{both}$ contains the embedding information in both branches, it does not lead to the same performance. This underlines the importance of the concatenation of the three features together compared to independently. 
Comparing pDeepFilterNet2$^{D}_{erb}$ and pDeepFilterNet2$^{D}_{df}$ allows one to understand the weight of each branch in the personalization process. 
We thus observe that concatenating the embedding in the DF branch only or in both branches leads to almost identical results, while concatenating it in the ERB branch only leads to lower results. 
We can guess that the embedding allow the DF branch to focus on removing everything but the target voice. In fact, isolating two voices is a difficult task which may explain why the DF branch has more impact than the ERB branch as it is used for fine estimation.

% \subsubsection{Computational complexity}
\noindent \textbf{Computational complexity.}\ 
We finally focus on the impact of personalization on the computational complexity of DeepFilterNet2. We compare the number of parameters, the multiply-accumulate operations (MAC) and the real-time factor (RTF) computed on a Intel® Core™ i7-10870H CPU @ 2.20GHz × 16.
We can see that simply concatenating the embedding within the unified encoder leads to minimal impact. 
Indeed, the MAC, the RTF and the number of parameters remain unchanged. 
Nonetheless, the three other personalization methods change the complexity of the framework. 
This was expected as deunifiying the encoder leads to an additional GRU that increases the complexity. 
In the end, even though the dual encoder architectures are slightly larger than the unified encoder architecture, the complexity is still very low compared to most of the recent PSE models.

\subsection{Blind test set}
We compare our best model on the generated test set, pDeepFilterNet2, to a larger model, the TEA-PSE 3.0 \cite{teapse3.0}. We compare them on the DNS5 blind test set using the DNSMOS \cite{reddy2021dnsmos}, and we also compare their computational complexity. 
We can first see that for both tracks, pDeepFilterNet2 is not as good as the TEA-PSE 3.0. 
Yet, the gap between both models is less important on Track 2 which corresponds to speakerphone excerpts.  This could be explained by the nature of our training set as it is closer to the speakerphone track.
Notably, the TEA-PSE 3.0 has 22.24 million trainable parameters and the number of MAC is 19.66G. Our model is therefore ten times smaller and the number of MAC fifty times lower than TEA-PSE 3.0.
Such a light model could be implemented on low-resource devices which is not the case for bigger models like the TEA-PSE 3.0.

\section{Conclusion \& Future work}
In this work, we proposed a robust method to personalize a lightweight state-of-the-art SE model. We explored the different possibilities to integrate the embedding into this dual-stage framework and we determined the best implementation. We showed that our personalization strategy greatly improves the performances in the presence of interfering speakers while preserving the low computational complexity of DeepFilterNet2 making it a great candidate for real-time PSE on embedded devices. 
Future work will look into improving the performances of our model to make it competitive with state-of-the-art SE metrics while further optimizing its complexity.

\newpage

\bibliographystyle{IEEEtran}
\bibliography{refs}

\begin{thebibliography}{10}
\providecommand{\url}[1]{#1}
\csname url@samestyle\endcsname
\providecommand{\newblock}{\relax}
\providecommand{\bibinfo}[2]{#2}
\providecommand{\BIBentrySTDinterwordspacing}{\spaceskip=0pt\relax}
\providecommand{\BIBentryALTinterwordstretchfactor}{4}
\providecommand{\BIBentryALTinterwordspacing}{\spaceskip=\fontdimen2\font plus
\BIBentryALTinterwordstretchfactor\fontdimen3\font minus \fontdimen4\font\relax}
\providecommand{\BIBforeignlanguage}[2]{{%
\expandafter\ifx\csname l@#1\endcsname\relax
\typeout{** WARNING: IEEEtran.bst: No hyphenation pattern has been}%
\typeout{** loaded for the language `#1'. Using the pattern for}%
\typeout{** the default language instead.}%
\else
\language=\csname l@#1\endcsname
\fi
#2}}
\providecommand{\BIBdecl}{\relax}
\BIBdecl

\bibitem{xu2020spex}
C.~Xu, W.~Rao, E.~S. Chng, and H.~Li, ``Spex: Multi-scale time domain speaker extraction network,'' \emph{IEEE Trans. Audio, Speech, Lang. Process.}, vol.~28, pp. 1370--1384, 2020.

\bibitem{ge2020spex+}
M.~Ge, C.~Xu, L.~Wang, E.~S. Chng, J.~Dang, and H.~Li, ``Spex+: A complete time domain speaker extraction network,'' \emph{Proc. Interspeech}, 2020.

\bibitem{ge2021spex++}
------, ``Multi-stage speaker extraction with utterance and frame-level reference signals,'' in \emph{Proc. ICASSP}, 2021, pp. 6109--6113.

\bibitem{ji2020speaker_joint}
X.~Ji, M.~Yu, C.~Zhang, D.~Su, T.~Yu, X.~Liu \emph{et~al.}, ``Speaker-aware target speaker enhancement by jointly learning with speaker embedding extraction,'' in \emph{Proc. ICASSP}.\hskip 1em plus 0.5em minus 0.4em\relax IEEE, 2020, pp. 7294--7298.

\bibitem{wang2018voicefilter}
Q.~Wang, H.~Muckenhirn, K.~Wilson, P.~Sridhar, Z.~Wu, J.~Hershey \emph{et~al.}, ``Voicefilter: Targeted voice separation by speaker-conditioned spectrogram masking,'' \emph{Proc. Interspeech}, 2018.

\bibitem{giri2021persopercepnet}
R.~Giri, S.~Venkataramani, J.-M. Valin, U.~Isik, and A.~Krishnaswamy, ``Personalized percepnet: Real-time, low-complexity target voice separation and enhancement,'' \emph{Proc. Interspeech}, 2021.

\bibitem{newmodels_pdccrn}
S.~E. Eskimez, T.~Yoshioka, H.~Wang, X.~Wang, Z.~Chen, and X.~Huang, ``Personalized speech enhancement: New models and comprehensive evaluation,'' in \emph{Proc. ICASSP}, 2022, pp. 356--360.

\bibitem{vzmolikova2019speakerbeam}
K.~{\v{Z}}mol{\'\i}kov{\'a}, M.~Delcroix, K.~Kinoshita, T.~Ochiai, T.~Nakatani, L.~Burget \emph{et~al.}, ``Speakerbeam: Speaker aware neural network for target speaker extraction in speech mixtures,'' \emph{IEEE Jour. of Selected Topics in Signal Processing}, vol.~13, no.~4, pp. 800--814, 2019.

\bibitem{2stage}
A.~Li, W.~Liu, X.~Luo, C.~Zheng, and X.~Li, ``Icassp 2021 deep noise suppression challenge: Decoupling magnitude and phase optimization with a two-stage deep network,'' in \emph{Proc. ICASSP}, 2021, pp. 6628--6632.

\bibitem{ju2022tea}
Y.~Ju, W.~Rao, X.~Yan, Y.~Fu, S.~Lv, L.~Cheng \emph{et~al.}, ``Tea-pse: Tencent-ethereal-audio-lab personalized speech enhancement system for icassp 2022 dns challenge,'' in \emph{Proc. ICASSP}.\hskip 1em plus 0.5em minus 0.4em\relax IEEE, 2022, pp. 9291--9295.

\bibitem{li2022glance}
A.~Li, C.~Zheng, L.~Zhang, and X.~Li, ``Glance and gaze: A collaborative learning framework for single-channel speech enhancement,'' \emph{Applied Acoustics}, vol. 187, p. 108499, 2022.

\bibitem{schroter2022deepfilternet2}
H.~Schr{\"o}ter, A.~Maier, A.~Escalante-B, and T.~Rosenkranz, ``Deepfilternet2: Towards real-time speech enhancement on embedded devices for full-band audio,'' in \emph{Proc. IWAENC}, 2022, pp. 1--5.

\bibitem{thakker2022fast}
M.~Thakker, S.~E. Eskimez, T.~Yoshioka, and H.~Wang, ``Fast real-time personalized speech enhancement: End-to-end enhancement network (e3net) and knowledge distillation,'' \emph{Proc. Interspeech}, 2022.

\bibitem{wang2020voicefilter}
Q.~Wang, I.~L. Moreno, M.~Saglam, K.~Wilson, A.~Chiao, R.~Liu \emph{et~al.}, ``Voicefilter-lite: Streaming targeted voice separation for on-device speech recognition,'' \emph{Proc. Interspeech}, 2020.

\bibitem{delcroix2019compact}
M.~Delcroix, K.~Zmolikova, T.~Ochiai, K.~Kinoshita, S.~Araki, and T.~Nakatani, ``Compact network for speakerbeam target speaker extraction,'' in \emph{Proc. ICASSP}, 2019, pp. 6965--6969.

\bibitem{ecapa}
B.~Desplanques, J.~Thienpondt, and K.~Demuynck, ``Ecapa-tdnn: Emphasized channel attention, propagation and aggregation in tdnn based speaker verification,'' \emph{Proc. Interspeech}, 2020.

\bibitem{snyder2018xvector}
D.~Snyder, D.~Garcia-Romero, G.~Sell, D.~Povey, and S.~Khudanpur, ``X-vectors: Robust dnn embeddings for speaker recognition,'' in \emph{Proc. ICASSP}, 2018, pp. 5329--5333.

\bibitem{osloss}
S.~E. Eskimez, T.~Yoshioka, H.~Wang, X.~Wang, Z.~Chen, and X.~Huang, ``Personalized speech enhancement: New models and comprehensive evaluation,'' in \emph{Proc. ICASSP}, 2022, pp. 356--360.

\bibitem{dubey2023icassp}
H.~Dubey, A.~Aazami, V.~Gopal, B.~Naderi, S.~Braun, R.~Cutler \emph{et~al.}, ``Icassp 2023 deep noise suppression challenge,'' \emph{Proc. ICASSP}, 2023.

\bibitem{ardila2019common}
R.~Ardila, M.~Branson, K.~Davis, M.~Henretty, M.~Kohler, J.~Meyer \emph{et~al.}, ``Common voice: A massively-multilingual speech corpus,'' \emph{LREC}, 2019.

\bibitem{wang2023upn}
Z.~Wang, R.~Giri, D.~Shah, J.-M. Valin, M.~M. Goodwin, and P.~Smaragdis, ``A framework for unified real-time personalized and non-personalized speech enhancement,'' in \emph{Proc. ICASSP}, 2023, pp. 1--5.

\bibitem{teapse3.0}
Y.~Ju, J.~Chen, S.~Zhang, S.~He, W.~Rao, W.~Zhu \emph{et~al.}, ``Tea-pse 3.0: Tencent-ethereal-audio-lab personalized speech enhancement system for icassp 2023 dns-challenge,'' in \emph{Proc. ICASSP}, 2023, pp. 1--2.

\bibitem{reddy2021dnsmos}
C.~K.~A. Reddy, V.~Gopal, and R.~Cutler, ``Dnsmos: A non-intrusive perceptual objective speech quality metric to evaluate noise suppressors,'' \emph{Proc. ICASSP}, 2021.

\end{thebibliography}

\end{document}